\documentclass[twocolumn,showpacs,prb,amsmath,amssymb,epsfig]{revtex4}

\usepackage{graphicx}
\usepackage{color}
\usepackage{dcolumn}
\usepackage{bm}

\begin{document}

\title{Local transport in a disorder-stabilized correlated
insulating phase}
\author{M. Baenninger, A. Ghosh, M. Pepper, H. E. Beere, I. Farrer,
P. Atkinson, D. A. Ritchie}
\affiliation{Cavendish Laboratory,
University of Cambridge, Madingley Road, Cambridge CB3 0HE, United
Kingdom}
\date{\today}

\begin{abstract}
We report the experimental realization of a correlated insulating
phase in 2D GaAs/AlGaAs heterostructures at low electron densities in
a limited window of background disorder. This has been achieved at
mesoscopic length scales, where the insulating phase is characterized
by a universal hopping transport mechanism. Transport in this regime
is determined only by the average electron separation, independent of
the topology of background disorder. We have discussed this
observation in terms of a pinned electron solid ground state,
stabilized by mutual interplay of disorder and Coulomb interaction.
\end{abstract}

 \maketitle

In the presence of Coulomb interaction, both magnetic field and
disorder are predicted to stabilize many-body charge-ordered ground
states.~\cite{phases,phase_B0} Strong perpendicular magnetic field
$B_\perp$ quenches the vibrational motion of electrons, and has been
extensively exploited to realize a charge-density wave (CDW) ground
state in systems with weak background disorder.~\cite{expt,micro}
Despite the effort however, the nature of localization in such
systems has been controversial, with both pinned Wigner solid (WS)
formation and inhomogeneity-driven percolation transition being
suggested.~\cite{shashkin} On the other hand, disorder stabilizes
Coulomb correlation effects by introducing a pinning gap $\Delta_{\rm
pin}$ in the phonon density of states, which provides a long
wavelength cutoff.~\cite{phase_B0} This has led to the theoretical
prediction of several forms of CDW ground states at zero or low
$B_\perp$. Systematic experimental investigations on such
possibilities, however, have been rare, and form the subject of this
work.

Increasing the magnitude of background potential fluctuations
increases $\Delta_{\rm pin}$ which stabilizes the CDW phases to
higher temperatures. In modulation-doped GaAs/AlGaAs
heterostructures, where disorder primarily arises from the charged
dopant ions,~\cite{efros} $\Delta_{\rm pin} \sim
\exp{(-4\pi\delta_{\rm sp}/\sqrt{3}r_{\rm ee})}$ depends strongly on
the setback distance $\delta_{\rm sp}$ that separates the 2D electron
system (2DES) and the dopants, where $r_{\rm ee}$ is the mean
distance between the electrons in the 2DES.~\cite{shklovskii1,chui}
However, disorder affects the ground state transport in two critical
ways. First, presence of $\Delta_{\rm pin}$ disintegrates the CDW
phase into domains of finite size $\lambda_{\rm d} \sim \mbox{sound
velocity}/\Delta_{\rm pin}$. At strong pinning, $\lambda_{\rm d}$
becomes microscopically small, leading to significant averaging in
transport measurements with conventional macroscopic devices. Second,
strong potential fluctuations can also result in a ``freezing'' of
transport below a certain percolation threshold even when electron
density ($n_{\rm s}$) is relatively high, thereby making the regime
of strong effective Coulomb interaction inaccessible.

Here, we show that these difficulties can be largely overcome by
using modulation-doped heterostructures of mesoscopic dimensions. In
such devices transport freezes at much lower $n_{\rm s}$ in
comparison to macroscopic devices at the same $\delta_{\rm sp}$ or
disorder, thereby allowing transport at a large interaction parameter
$r_{\rm s} = 1/a_{\rm B}^*\sqrt{\pi n_{\rm s}} \sim 7 - 8$ ($a_{\rm
B}^*$ is the effective Bohr radius), even when $\delta_{\rm sp}$ is
relatively small. Typical dimension $L$ of our devices in the current
carrying direction was chosen to be $\sim 2 - 4 \mu$m, which is also
similar in order of magnitude to the $\lambda_{\rm d}$ suggested by
recent microwave absorption studies for pinned WS ground
states.~\cite{micro} The low-$B_\perp$ magnetotransport in these
devices was found to display a striking universality in that the
hopping distance in the localized regime was determined by $r_{\rm
ee} = 1/\sqrt{n_{\rm s}}$, rather than the details of background
disorder, indicating an unusual self-localization of electrons at
sufficiently low $n_{\rm s}$.

\begin{table}[b]
\caption{\label{tab:table1} Geometrical and structural property of
the devices. $n_\delta$ is the density of Si dopants, and $W$ is the
width. The background doping concentration is $\lesssim
10^{14}$cm$^{-3}$ in all devices. }
\begin{ruledtabular}
\begin{tabular}{llcccl}
Wafer&Device&$\delta_{\rm sp}$&$n_{\delta}$&$W\times L$& Doping\\
     &      &   nm            & $10^{12}$cm$^{-2}$   &
     $\mu$m$\times\mu$m &\\
\hline
A2407 & A07a & 20 & 2.5 & $8\times 2$ & $\delta$\\
    & A07b & 20 & 2.5 & $8\times 3$ & $\delta$\\
A2678 & A78a & 40 & 2.5 & $8\times 2.5$ & $\delta$\\
    & A78b & 40 & 2.5 & $8\times 4$ & $\delta$\\
A2677 & A77 & 40 & $-$\footnotemark[1] & $8\times 3$ & Bulk\\
    & A77L & 40 & $-$\footnotemark[1] & $100\times900$ & Bulk\\
C2367 & C67 & 60 & 0.7 & $8\times 3$ & $\delta$\\
T546 & T46 & 80 & 1.9 & $8\times 3$ & $\delta$\\
\end{tabular}
\end{ruledtabular}
\footnotetext[1]{The doping concentration of bulk doped devices is
$2\times 10^{18}$cm$^{-3}$ over a range of 40nm.}
\end{table}

We have used Si modulation-doped GaAs/AlGaAs heterostructures where
$\delta_{\rm sp}$ was varied from 20 nm to 80 nm. At a fixed $n_{\rm
s}$, the effect of $\delta_{\rm sp}$ on the strength of potential
fluctuations is reflected in the mobility $\mu$, as can be observed
from Fig.~1b. Both monolayer ($\delta$)- and bulk-doped wafers were
used. Relevant properties of the devices are given in Table I.
Devices were cooled from room temperature to 4.2 K over 24 - 36 hours
to allow maximal correlation in the dopant layer (redistribution of
DX-centers).~\cite{dopants} This slow cooldown technique also lead to
excellent reproducibility over repeated thermal cycles. Electrical
measurements were carried out with standard low frequency (7.2 Hz)
four-probe technique with excitation current of $\sim 0.01 - 0.1$ nA
to minimize heating and other nonlinear effects. A direct measurement
of $n_{\rm s}$ within the microscopic region was carried out with an
edge-state reflection-based technique.~\cite{self1}

\begin{figure}[t]
\centering
\includegraphics[height=9.326cm,width=7.5cm]{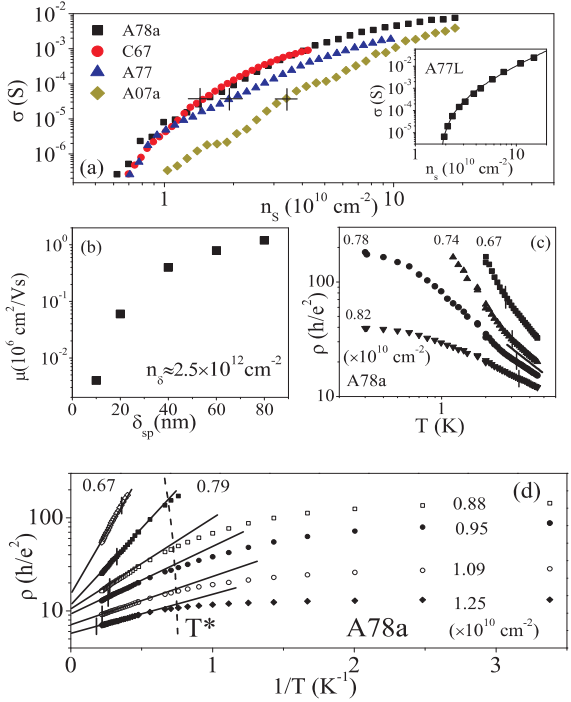}
\caption{(Color online) (a) Conductivity ($\sigma$) of mesoscopic
samples as a function of electron density $n_{\rm s}$ at $T
\approx 0.3$ K. The crosses denote $n_{\rm s}^*$ for individual
samples (see text). Inset: $n_{\rm s}$-dependence of $\sigma$ for
a macroscopic Hall bar A77L. The solid line is the best fit of a
classical percolation-like scaling relation $\sigma \sim (n_{\rm
s} - n_{\rm c})^\gamma$. (b) $\delta_{\rm sp}$ dependence of
mobility at constant $n_{\rm s}$ and $n_{\delta}$ for
heterostructures similar to those used in presented work (c)
Resistivity ($\rho$) as a function of temperature measured at
$B_\perp = 1$ T. The solid line represent a power law of $\sim
T^{-1}$, vertical lines in (c) and (d) indicate the Fermi
temperatures $T_{\rm F}$. (d) Activation and saturation of $\rho$
at $B_\perp = 1.5$ T.}
\end{figure}

In Fig.~1a we compare the $n_{\rm s}$-scale of localization
transition at $B_\perp = 0$ and $T = 0.3$ K in macroscopic and
microscopic devices from the same wafer. In a standard $100
\mu$m$\times 900 \mu$m Hall bar, as illustrated with wafer A2677, the
linear conductivity $\sigma \rightarrow 0$ (A77L: Inset of Fig.~1a)
at $\sim 3$ times the $n_{\rm s}$ compared to the mesoscopic sample
(A77) from the same wafer. Further, $\sigma$ in the large sample A77L
shows excellent classical percolation-like scaling $\sigma \sim
(n_{\rm s} - n_{\rm c})^\gamma$ ($n_{\rm c} = 1.72\times10^{10}$
cm$^{-2}$), where $\gamma \approx 2$, implying a inhomogeneity-driven
percolation transition at nonzero $T$~\cite{shashkin} (solid line in
inset of Fig.~1a). Similar scaling in the mesoscopic systems,
however, was found to be difficult, with unphysically large estimates
of $\gamma \sim 3.2 - 3.7$ (not shown), indicating a different
mechanism of localization transition.

As $n_{\rm s}$ is lowered below a sample-dependent characteristic
scale $n_{\rm s}^*$ (denoted by the crosses in Fig.~1a), onset of
strong localization is identified by the resistivity $\rho$ ($=
1/\sigma$) exceeding $\approx h/e^2$. At $n_{\rm s} \ll n_{\rm s}^*$,
the $T$-dependence of $\rho$ at a fixed $n_{\rm s}$ can be divided
into three regimes, as illustrated with A78a: First, transport in the
classical regime at $T \gtrsim T_{\rm F}$ is magnified in Fig.~1c,
where $T_{\rm F}$ is the Fermi temperature. In this regime $\rho
\propto T^{-\beta}$, where $\beta \sim 1$ (indicated by the solid
line). As $T$ is decreased, the onset of the quantum regime ($T
\lesssim T_{\rm F}$) results in stronger increase in $\rho$ with
decreasing $T$. Note that the clear classical to quantum crossover
implies a well-defined $T_{\rm F}$, and hence a uniform charge
density distribution down to the lowest $n_{\rm s} \sim
6.5\times10^{9}$ cm$^{-2}$ (In A77L, inhomogeneity sets in at $n_{\rm
s}$ as large as $\sim 4-5\times 10^{10}$ cm$^{-2}$). In the quantum
regime and for $T_{\rm F} > T > T^*$, Fig.~1d shows that the behavior
of $\rho$ is activated with $\rho(T) = \rho_3\exp(\epsilon_3/k_{\rm
B}T)$, where $\epsilon_3$ is the activation energy. From the $n_{\rm
s}$- and $B_\perp$-dependence of the pre-exponential $\rho_3$, we
have shown earlier that the transport mechanism in this regime
corresponds to nearest-neighbor hopping.~\cite{self1} Below the
characteristic scale $T^* \sim 1$ K, variation of $\rho$ becomes
weak, tending to a finite magnitude even in the strongly localized
regime. This saturation in the insulating regime cannot be explained
in terms of an elevated electron temperature due to insufficient
thermal coupling to the lattice since $T^*$ depends only weakly on
electron density up to $n_{\rm s} \sim n_{\rm s}^*$ (Fig.~1d), and
the damping of Shubnikov-de Haas oscillations in the metallic regime
shows the base electron temperature to be $\approx 300$ mK.

\begin{figure}[t]
\centering
\includegraphics[height=6.274cm,width=7cm]{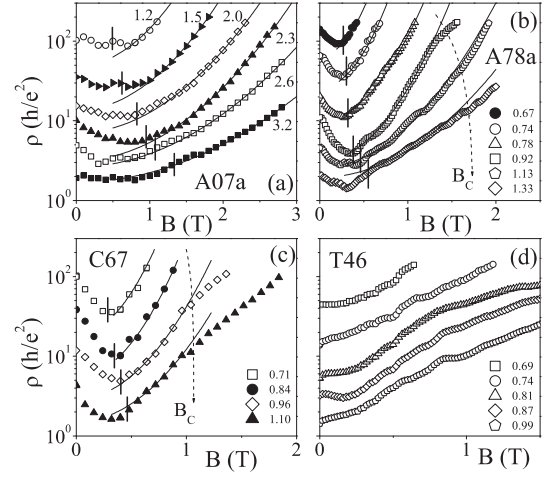}
\caption{Typical magnetoresistivity traces in four samples with
varying level of disorder. The vertical lines denote $\nu = 1$.
The numbers indicate electron density in units of $10^{10}$
cm$^{-2}$. $B_{\rm c}$ denotes the field scale up to which a
quadratic $B_\perp$-dependence could be observed. The parameters
$\alpha$ and $\rho_{\rm B}$ were obtained from the slope and
y-intercept of linear fits to $\ln(\rho)-B_\perp^2$traces,
respectively.}
\end{figure}

In order to explore the physical mechanism behind the weak
$T$-dependence of $\rho$, we have carried out extensive
magnetoresistivity (MR) measurements at the base $T$. Figs.~2a to ~2d
show the $B_\perp$-dependence of MR in the insulating regime of four
devices with increasing $\delta_{\rm sp}$ from 20 nm to 80 nm. In
general, we find a strong negative MR in A07, A78 and C67 at low
$B_\perp$, which can be attributed to interference of hopping paths.
The negative MR is followed by an exponential rise in $\rho$ as
$B_\perp$ is increased further. We have recently shown that the
logarithm of such a positive MR at low $B_\perp$ varies in a
quadratic manner with $B_\perp$, i.e., $\rho(B_\perp) = \rho_{\rm
B}\exp(\alpha B_\perp^2)$, where $\rho_{\rm B}$ and $\alpha$ are
$n_{\rm s}$-dependent factors.~\cite{self1} Such a variation, denoted
by the solid lines in Fig.~2, is found to be limited to $n_{\rm s}
\lesssim n_{\rm s}^*$, and extends over a $B_\perp$-scale of $B_{\rm
c}$, where $B_{\rm c}$ was found to decrease rapidly as $\delta_{\rm
sp}$ is increased. Note that in T46 (lowest disorder), neither a
clear negative MR nor an exponential $B_\perp^2$-dependence were
observed. A physical significance of $B_{\rm c}$ and of the
qualitatively different MR behavior of T46 will be discussed later.

\begin{figure}[t]
\centering
\includegraphics[height=5.609cm,width=7cm]{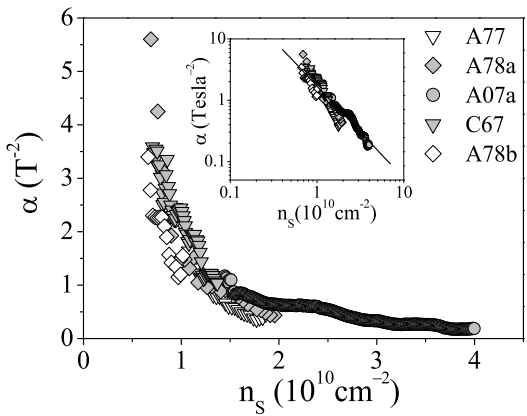}
\caption{Absolute magnitude of $\alpha$ obtained from the slope of
$\ln(\rho)-B_\perp^2$ traces for five different samples. Inset shows
the same data in a $\log$-$\log$ scale. The slope of the solid line
is $-3/2$.}
\end{figure}

The observed behavior of $\rho$ can be naturally explained in the
framework of tunnelling of electrons between two trap sites separated
by a distance $r_{\rm ij}$. In weak $B_\perp$, such that the magnetic
length $\lambda = \sqrt{\hbar/eB_\perp} \gg \xi$, where $\xi$ is the
localization length, the asymptotic form of the hydrogenic wave
function changes from $\psi(r) \sim \exp(-r/\xi)$ to $\psi(r) \sim
\exp(-r/\xi - r^3\xi/24\lambda^4)$.~\cite{shklovskii} This leads to a
MR, $\rho(B_\perp) = \rho_0 \exp(2r_{\rm ij}/\xi)\exp(Ce^2r_{\rm
ij}^3\xi B_\perp^2/12\hbar^2)$, which implies

\begin{equation}
\label{alpha} \rho_{\rm B} = \rho_0\exp(2r_{\rm ij}/\xi) \quad
\mbox{and} \quad \alpha = Ce^2r_{\rm ij}^3\xi/12\hbar^2.
\end{equation}

\noindent While $\rho_{\rm B}$ depends on the tunnelling probability
at $B_\perp = 0$, $\alpha$ denotes the rate of change of this
probability when $B_\perp$ is switched on. Importantly, both
parameters provide information on the intersite distance $r_{\rm
ij}$, as well as $\xi$ independently. The parameter $C \sim 0.5 - 1$
depends on the number of bonds at percolation threshold in the random
resistor network (we shall subsequently assume $C \approx 1$). Since
conventional hopping sites are essentially impurity states, both
$\alpha$ and $\rho_{\rm B}$ are expected to be strongly disorder
dependent. Note that, since wave function overlap plays a critical
role in transport, a direct source-to-drain tunnelling is ruled out
in our case.~\cite{savchenko}

From the MR data we have evaluated $\alpha$ and $\rho_{\rm B}$
from the slope and intercept of the $\ln(\rho)-B_\perp^2$ traces.
Further details of the analysis can be found
elsewhere.~\cite{self1} In Fig.~3 we have shown $\alpha$ as a
function of $n_{\rm s}$ for five different samples up to the
corresponding $n_{\rm s}^*$. Strikingly, the absolute magnitudes
of $\alpha$ from different samples are strongly correlated, and
can be described by an universal $n_{\rm s}$-dependent function
over nearly two orders of magnitude. At stronger disorder (e.g.,
A07), localization occurs at a higher $n_{\rm s}$ resulting in a
lower $\alpha$, while at lower disorder (e.g. C67) localization
occurs at lower $n_{\rm s}$ yielding a larger magnitude of
$\alpha$. This indicates that magnetotransport in such mesoscopic
samples is not determined directly by disorder, but by $n_{\rm s}$
in the localized regime. Qualitatively, the decreasing behavior of
$\alpha$ with increasing $n_{\rm s}$ itself is inconsistent with
the single-particle localization in an Anderson
insulator.~\cite{self1,fowler2}

From the strong sample-to-sample correlation in the magnitude of
$\alpha$, a disorder-associated origin of $r_{\rm ij}$, is clearly
unlikely. For example, taking $r_{\rm ij} \sim \delta_{\rm sp}$ will
lead to distinct sets of $\alpha$ for wafers with different
$\delta_{\rm sp}$. However, in the context of a pinned CDW ground
state, another relevant length scale is $r_{\rm ee}$. Indeed, in case
of tunnelling events over a mean electron separation, i.e., $r_{\rm
ij} \approx r_{\rm ee}$, we find that Eq.~\ref{alpha} describes both
absolute magnitude, as well as the $n_{\rm s}$-dependence of $\alpha$
quantitatively. Using $r_{\rm ij} \approx 1/\sqrt{n_{\rm s}}$,
Eq.~\ref{alpha} leads to $\alpha \propto n_{\rm s}^{-3/2}$, as indeed
observed experimentally (solid line in the inset of Fig.~3). Allowing
for sample-to-sample variation, we find $\alpha =
(1.7\pm0.5)\times10^{21}/n_{\rm s}^{3/2}$ T$^{-2}$ from which, using
Eq.~\ref{alpha}, we get $\xi = 9.0\pm2.6$ nm, which is close to
$a_{\rm B}^*$ in GaAs ($\approx 10.5$ nm).

The analysis can be immediately checked for consistency from the
$r_{\rm ee}$-dependence of $\rho_{\rm B}$. From Fig.~4, we find that
$\rho_{\rm B}$ increases strongly with increasing $r_{\rm ee}$ when
$n_{\rm s} \ll n_{\rm s}^*$, as expected in the simple tunnelling
framework (Eq.~\ref{alpha}). In spite of the scatter, the overall
slopes of the $\ln(\rho_{\rm B}) - r_{\rm ee}$ plots are similar in
different samples (solid lines) with $\xi$ estimated to be $\approx
13\pm4$ nm, agreeing with that obtained from the analysis of
$\alpha$. Note that the $\rho_{\rm B}$ deviates from the exponential
dependence, and tends to saturate as $n_{\rm s} \rightarrow n_{\rm
s}^*$. While this is not completely understood at present, we note
that the saturation in $\rho_{\rm B}$ occurs within the range
$\rho_{\rm B} \sim 1 - 2\times h/e^2$, irrespective of sample
details. Similar universality in the hopping pre-exponential has been
observed in the context of $T$-dependence of $\rho$ in variable-range
hopping,~\cite{khondaker} and has been suggested to indicate an
electron-electron interaction mediated energy-transfer mechanism.

\begin{figure}[t]
\centering
\includegraphics[height=4.147cm,width=7cm]{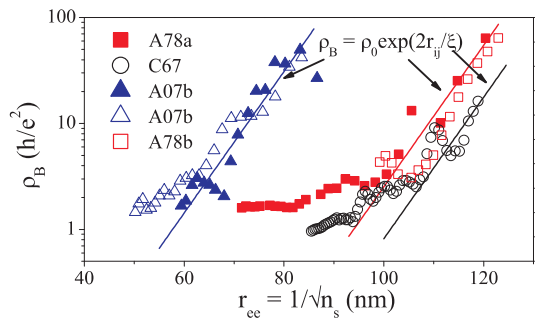}
\caption{(Color online) The dependence of $\rho_{\rm B}$ on the
average electron separation $r_{\rm ee}$ in five different
samples. The slope of the solid lines gives an estimate of $\xi$
(Eq.~\ref{alpha}).}
\end{figure}

We now discuss the physical scenario which could lead to the electron
separation-dependent hopping transport. We show that our observations
can be explained in the theoretical framework of defect motion in a
quantum solid that was originally developed by Andreev and Lifshitz
in context of solid He$^3$,~\cite{AL} and later adapted for a WS
ground state.~\cite{spivak,pichard} In our case, transport in both
quantum and classical regime can be understood in terms of tunnelling
of localized defects in an interaction-induced pinned electron solid
phase as $n_{\rm s}$ is reduced below the melting point $n_{\rm
s}^*$. The defects, which act as quasiparticles at low $T$, can arise
from regular interstitials, vacancies, dislocation loops etc., as
well as from zero-point vibration of individual lattice
sites.~\cite{AL} The scale of zero-point fluctuation $\sim h/r_{\rm
ee}\sqrt{m^*U_{\rm C}} \approx 2\pi/\sqrt{r_{\rm s}} \gtrsim 1$, is
indeed strong in our case over the experimental range of $n_{\rm s}$,
where $U_{\rm C} \approx e^2/4\pi\epsilon r_{\rm ee}$ is the
interatomic interaction energy scale.

In the quantum regime, the transport at higher $T$ ($T_{\rm F} \gg
T \gg T^*$) is predicted to be thermally activated
nearest-neighbor hopping of localized defects, while at lower $T$
($\ll T^*$) tunnelling of such defects leads to a $T$-independent
transport.~\cite{AL} While this clearly describes the weak
$T$-dependence of $\rho$ at low temperatures (Fig.~1d), the
strongest support to this picture comes from the fact that the
natural length scale of tunnelling is indeed the average electron
separation $r_{\rm ee}$. This immediately explains the unusual
$n_{\rm s}$ (or $r_{\rm ee}$)-dependence of both $\alpha$ and
$\rho_{\rm B}$, as well as the apparent insensitivity of these
parameters to local disorder. The negative MR at low $B_{\perp}$
caused by destruction of interference is then expected to persist
up to a $B_{\perp}$ corresponding to $\nu = n_{\rm s}h/eB_\perp
\sim 1$ (one flux quantum $\phi_0$ within an area of $r_{\rm
ee}^2$), as indeed observed in our experiments (Fig.~2). The
tunnelling of defect scenario also allows an estimate of the
crossover scale $k_{\rm B}T^*=\epsilon_3/\ln(\Delta_{\rm
pin}/\Delta\epsilon)$,~\cite{AL} where $\Delta\epsilon$ is the
bandwidth. For a pinned WS ground state, using the expression of
$\Delta_{\rm pin}$ in Ref.[8], experimentally measured
$\epsilon_3$, and $\Delta\epsilon \sim h^2/8m^*r_{\rm ee}^2$, we
find $T^* \sim$ \textsl{O}[1 Kelvin] over the experimental range
of $n_{\rm s}$ in A78a, giving good order-of-magnitude agreement
to the observed scale of $T^*$. Finally, the behavior of $\rho
\sim T^{-1}$ in the classical regime ($T > T_{\rm F}$)(Fig.~1c)
has also been recently observed,~\cite{lilly} and interpreted in
terms of transport mediated by defect-type topological objects
(Fermi-liquid droplets) in the WS phase.~\cite{spivak}

In presence of pinning, the MR-data suggests the asymptotic form of
the wave function $\psi(r) \sim \exp(r/\xi)$, where $\xi \approx
a_{\rm B}^*$. However, the interplay of confinement arising from the
magnetic potential and disorder pinning is expected to be critical in
determining $\psi(r)$, with disorder pinning dominating at low
$B_\perp$. This is expected to result in the upper cutoff $B_{\rm c}$
that decreases with decreasing disorder, as observed experimentally.
The intricate interplay between disorder, electron-electron
interaction and magnetic field is further illustrated by the absence
of a clear $B_{\perp}^2$ dependence of the MR in T46 (largest
$\delta_{\rm sp}$), which could be explained by a prohibitively small
$B_{\rm c}$ or the very instability of the solid phase at
sufficiently low disorder. On the other hand, devices with
$\delta_{\rm sp} \lesssim 10$nm showed inhomogeneity-driven Coulomb
blockade oscillations in the localized regime, making the
investigation of such a charge correlated state impossible. A
quantitative understanding of the scale of $B_{\rm c}$, as well as
the specific spatial structure of the ground state in the
intermediate disorder regime, will require further investigations,
which are presently in progress.

\end{document}